\documentclass[%
 reprint,
 amsmath,amssymb,
 aps,
]{revtex4-2}

\usepackage{graphicx}
\usepackage{dcolumn}
\usepackage{bm}
\usepackage{gensymb}
\usepackage[usenames,dvipsnames]{color} 
\usepackage[caption=false]{subfig}
\usepackage{url}
\usepackage{hyperref}
\usepackage[mathlines]{lineno}
\newcommand{\specificthanks}[1]{\@fnsymbol{#1}}

\begin{document}

\preprint{}

\title{High voltage DC gun for high intensity polarized electron source}

\author{Erdong Wang} 
\email{Corresponding author\\E-mail: wange@bnl.gov}
\thanks{\\These authors contributed equally to this study}
\author{Omer Rahman}
\thanks{These authors contributed equally to this study}
\affiliation{Brookhaven National Laboratory, Upton, NY 11973,USA}
\author{John Skaritka}
\affiliation{Brookhaven National Laboratory, Upton, NY 11973,USA}
\author{Wei Liu}
\affiliation{Brookhaven National Laboratory, Upton, NY 11973,USA}
\author{Jyoti Biswas}
\affiliation{Brookhaven National Laboratory, Upton, NY 11973,USA}
\author{Chrisopher Degen}
\affiliation{Brookhaven National Laboratory, Upton, NY 11973,USA}
\author{Patrick Inacker}
\affiliation{Brookhaven National Laboratory, Upton, NY 11973,USA}
\author{Robert Lambiase}
\affiliation{Brookhaven National Laboratory, Upton, NY 11973,USA}
\author{Matthew Paniccia}
\affiliation{Brookhaven National Laboratory, Upton, NY 11973,USA}


\date{\today}

\begin{abstract}
The high intensity polarized electron source is a critical component for future nuclear physics facilities. The Electron Ion Collider (EIC) requires a polarized electron gun with higher voltage and higher bunch charge compared to any existing polarized electron source. At Brookhaven National Laboratory, we have built an inverted high voltage direct current (HVDC) photoemission gun with a large cathode size. We report on the performances of GaAs photocathodes in a high gradient with up to a 16 nC bunch charge. The measurements were performed at a stable operating gap voltage of 300 kV - demonstrating outstanding lifetime, and robustness. We observed obvious lifetime enhancement by biasing the anode. The gun also integrated a cathode cooling system for potential application on high current electron sources. The various novel features implemented and demonstrated in this polarized HVDC gun open the door towards future high intensity-high average current electron accelerator facilities. 
\end{abstract}

\maketitle


\section{Introduction}

The Electron-Ion Collider (EIC) at Brookhaven National Laboratory(BNL) will be colliding polarized electrons with polarized hadrons to probe into various unsolved mysteries of Nuclear Physics \cite{EICwhite}. In order to achieve the expected luminosity of the collider, a high bunch charge, high intensity polarized electron source is required \cite{eic_cdr}. The polarized electron source should be able to provide 5.5-7 nC per bunch in a bunch train structure with 8 bunches every second. To achieve the required beam qualities before injecting the beam into the Rapid Cycling Synchrotron (RCS) and minimize beam loss, the gun voltage has to be above 280 kV with beam peak current above 4.5 A according to the current pre-injector design \cite{IPAC_PI}. \par
There have been substantial efforts in high voltage polarized gun development in the last few decades. The Stanford Linear Collider (SLC) gun has demonstrated extraction of up to 12 nC bunch charge at 120 Hz . The gap voltage for the SLC gun was 120 kV and the cathode had to be reactivated every 3-5 days \cite{clendenin1995polarized}. JLab Continuous Electron Beam Accelerator Facility (CEBAF) gun operates at $<$ 10 pC bunch charge with gap voltage up to 200 kV. Cathode operational lifetime at CEBAF is above one month in the operation of 200 uA average beam current \cite{PhysRevSTAB.10.023501}. Until now, all the polarized guns have operated at a gap voltage less than 200 kV to eliminate field emission\cite{grame2007,clendenin1995polarized,BastaniNejad2011a,Schultz1994,ELSA2003,TSENTALOVICH2007413,Nagoya2003,NIKHEF98}. On the peak current aspect, SLC and Nagoya DC guns have achieved 2-6 A of peak current. However, the rest of polarized guns have peak currents up to 10s of mA range. So far, no operational or retired polarized gun can fully meet the EIC polarized electron source requirements.\par
 At BNL, we have developed a high intensity polarized electron gun based on an inverted high voltage feedthrough design with a higher gap voltage and large cathode size. This gun can consistently operate at 300 kV gap voltage without any detectable field emission. The following features have been implemented into the gun to generate high intensity polarized electron beam: 1) Inverted HV feedthrough to minimize outgassing area; 2) Ceramic feedthrough and two triple-point sheds to minimize the electric field gradient at the ceramic-vacuum-metal joint and linearize the voltage along the length of the ceramic; 3) Movable and electrically insulated anode; 4) A HV cable with a slightly conductive jacket, which prevents charge build up, but does not deliver energy to an arc as a conductive shield would. Moreover, for future high current applications, we have developed a cathode cooling system integrated into the high voltage feedthrough in the gun. 
 In this article, we report on our HVDC gun design, gun performance including high voltage and vacuum performance, space charge limit, and lifetime measurements with 7.5 nC bunch charge.

\section{Inverted HVDC gun design considerations}

For an HVDC gun to deliver high intensity polarized electron beam with a good operational lifetime, a few considerations are critical during the inception stage. These considerations are implemented to prolong charge lifetime by minimizing cathode quantum efficiency (QE) decay while delivering high intensity beam with good quality and minimal loss.
The cathode QE is determined by the photon energy $h \nu$, material Density of State (DoS), and surface electron affinity ($E_a$). GaAs Photocathode, once activated, is a negative electron affinity($E_a<$ 0) photocathode. When in threshold photoemission, photon energy $h \nu \approx Eg$ ($E_g$ = bandgap energy), any small variation in $E_a$ could affect the QE dramatically. To produce high polarization electron beam, the cathode has to be operated in threshold photoemission mode which results in QE being very sensitive to the degradation of the activation layer which usually will increase $E_a$. To extend cathode long lifetime, the following effects have to be considered while designing the gun:
\begin{enumerate}
\item Ultra-high vacuum: The GaAs cathode emission surface is coated by a O(NF3)-Cs which is very sensitive to gas contamination including carbohydrate, oxygen, water, and other active gases. The active gases degrade the quantum efficiency (QE) by chemical poisoning or via ion back bombardment. Thus to obtain a long lifetime, the gun chamber dynamic pressure should be in $10^{-12}$ Torr scale with hydrogen being the dominant gas, above 99\% of the gas load.
\item High voltage: For high bunch charge beam, high gradient and high voltage are desirable on the cathode. However, the gradient cannot exceed 10 MV/m to avoid field emission. Depending on the desired bunch charge, the cathode size could be determined so that an adequately sized laser spot can be used to avoid the space charge limit and surface charge limit. To avoid electrons punching through the ceramic, an appropriate shed should be designed to protect the ceramic feedthrough. 
\item Beam loss: The beam loss close to the gun will cause an increase in the dynamic pressure. The main source of beam loss is from the beam halo. Therefore, a large beam pipe at the exit of the gun is necessary to minimize the effect of beam loss.
\item Ion Back Bombardment: Ion back bombardment is a mechanism that limits the cathode lifetime. Although ion back bombardment from ions generated in the DC gap cannot be eliminated, a biased anode is very useful in eliminating ions that are generated downstream \cite{grame08,Pozdeyev07,omer19}.
\item Masking of cathode active area: Electrons emitted from the edge of the cathode can take extreme trajectories due to strong transverse kick. Also, electron beam halo caused by the laser transverse halo can end up hitting the wall creating outgassing.  Thus, the emission area should be kept at a safe distance from the edge of the cathode. 
\end{enumerate}
We chose the load-lock inverted feedthrough as our gun type. Since one side of the inverted HV feedthrough is in a vacuum and another side is in contact with silicone grease and rubber, it minimizes the creepage compared to the case when the ceramic is exposed to the atmosphere. Therefore, the ceramic feedthrough size can be reduced which will result in a compact chamber design. This structure also eliminated the field emission electrons hitting the ceramic feedthrough \cite{carlos19}. The compact chamber reduces the outgassing surface area and results in ultra-high vacuum in the chamber with sufficient pumping. With the cathode load-lock system, the cathode can be activated at a separate preparation chamber, other than activating in the gun chamber such as SLC gun or JLab-FEL HVDC gun \cite{clendenin1995polarized,JLABFELSIGGINS2001}. A cathode load lock system is used to permit the exchange of the photocathode within an hour. \par 


A large cathode with a sufficiently large laser spot size can generate a high peak current electron beam. However, too large of a cathode size will drastically increase the overall cost of the vacuum components. For this gun, the cathode size was designed to be 1.26 cm in radius so that the cathode puck can be transferred through a 4.5 inch all-metal valve.
 The bunch charge of 7 nC cannot be a pancake shape from a DC gun which has a gradient on the cathode to be less than 5 MV/m. The space charge limit for the pencil shape beam is 
\begin{equation}
    J_{2d}=2.33 \times 10^{-6} V^{3/2}/d(1+\frac{d}{4r})
\end{equation}
where V is the voltage across the DC gap of width d, and r is the laser spot size radius \cite{scl_PRL}. The gap distance is determined by the maximum gradient on the electrode which has to be less than 10 MV/m to avoid field emission. The EIC pre-injector baseline design shows that to achieve 7 nC bunch charge from a 300 kV gun with minimum emittance and energy spread, the peak current is 4.5 A, the FWHM bunch length is 1.6 ns, and beam size is 8 mm \cite{eic_cdr,IPAC_PI}. To get minimal required beam quality before injecting beam into the RCS, the gun operational voltage has to be above 280 kV.
The commercially available inverted feedthrough R30 can hold high voltage up to 400 kV \cite{sctweb}. To be conservative, we designed our gun voltage up to 350 kV. Having higher voltage has the following advantages: 1) Can generate high charge beam with high peak current by increasing the space charge limit; 2) Eliminate the surface charge limit by lowering the surface barrier\cite{MULHOLLAN2001309,MARUYAMA2002199} 3) Preserve the beam longitudinal and transverse emittance during ballistic compression in the injector.\par

\begin{table}
\caption{DC gun design parameters }
\begin{ruledtabular}
\begin{tabular}{lcr}
& Value\\
    \hline
         Gap Voltage & 350 kV  \\
         Cathode radius & 1.26 cm  \\
         Pierce angle & 22 $\degree$ \\
         Cathode gradient & 3.8 MV/m \\
         Maximum gradient on the electrodes & 9.4 $\rm MV/m$ \\
         Anode radius & 1.8 cm\\
         Peak Current & 4.5 A\\
         Bunch charge & 7 nC \\
         Normalized Emittance  & 3.4 mm-mrad
\end{tabular}
\end{ruledtabular}
\label{tab:gun_parameter}
\end{table}

\begin{figure}[h]
\centering
\includegraphics[width=0.45\textwidth]{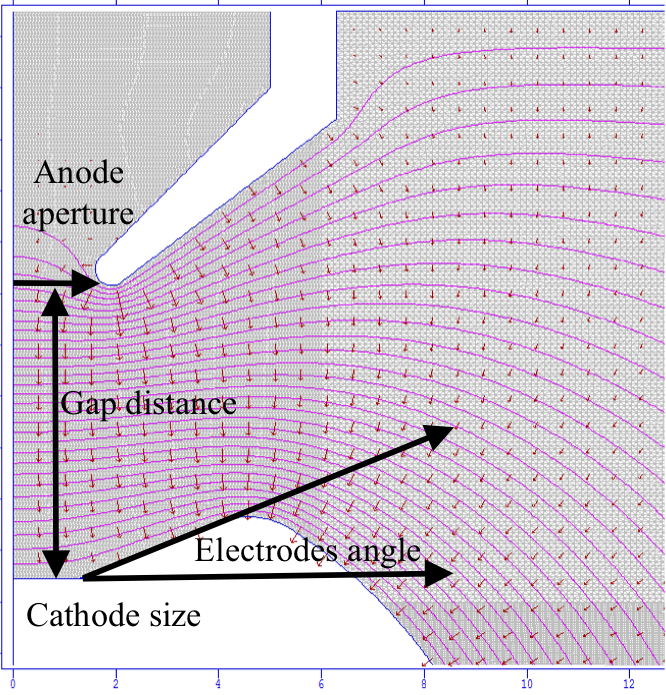}
\caption{DC gap geometry.}
\label{fig:gun_geo}
\end{figure}

The DC gun Pierce geometry was optimized by achieving small emittance after space charge compensation downstream of the 1st solenoid. The optimal parameters are the
cathode size ($L_c$), Pierce angle($\alpha_p$), gap distance, and the ratio of Pierce focusing size to the cathode size as shown in figure \ref{fig:gun_geo}.  Then, the anode aperture size is increased up to 1.4 times of the cathode size, to make sure that the stray beam from laser transverse halo or scattered photons can be extracted from the gun without any loss on the anode. The gun geometry parameters and simulated beam quality parameters are listed in Table ~\ref{tab:gun_parameter}.

 A bulk GaAs wafer was installed in a Molybdenum puck with a tungsten cap covering the edge of the wafer. The GaAs wafers were activated in a preparation chamber that is connected to the gun through the cathode storage chamber.  The cathodes were atomically cleaned by heat treatment up to $580 \degree $C with a  temperature ramping rate of $3 \degree $C/min. After the cathode has reached room temperature, successive deposition of Cs and O$_2$ can generate a typical QE of about 6\% at 785 nm. The activated cathodes were transferred to the storage chamber for storage and then eventually transferred to the gun for beam extraction.

\section{High voltage performance}

Once the electrode and anode geometries were optimized for best beam quality, the next considerations are the following components: high voltage feedthrough,triple-point sheds, anode shape, and NEG pump positioning. We used a 90 cm diameter spherical chamber as the gun chamber, with 20 cm in diameter spherical electrode to reduce the surface gradient. The HV feedthrough used for this gun is carbon-doped alumina which allows trapped electrons on the ceramic surface to be removed \cite{sct-insu}. 

\begin{figure}[!htb]
\centering
    \subfloat[\label{fig:gun_map}The field gradient map on the DC gap.]
    {\includegraphics[width=1\columnwidth]{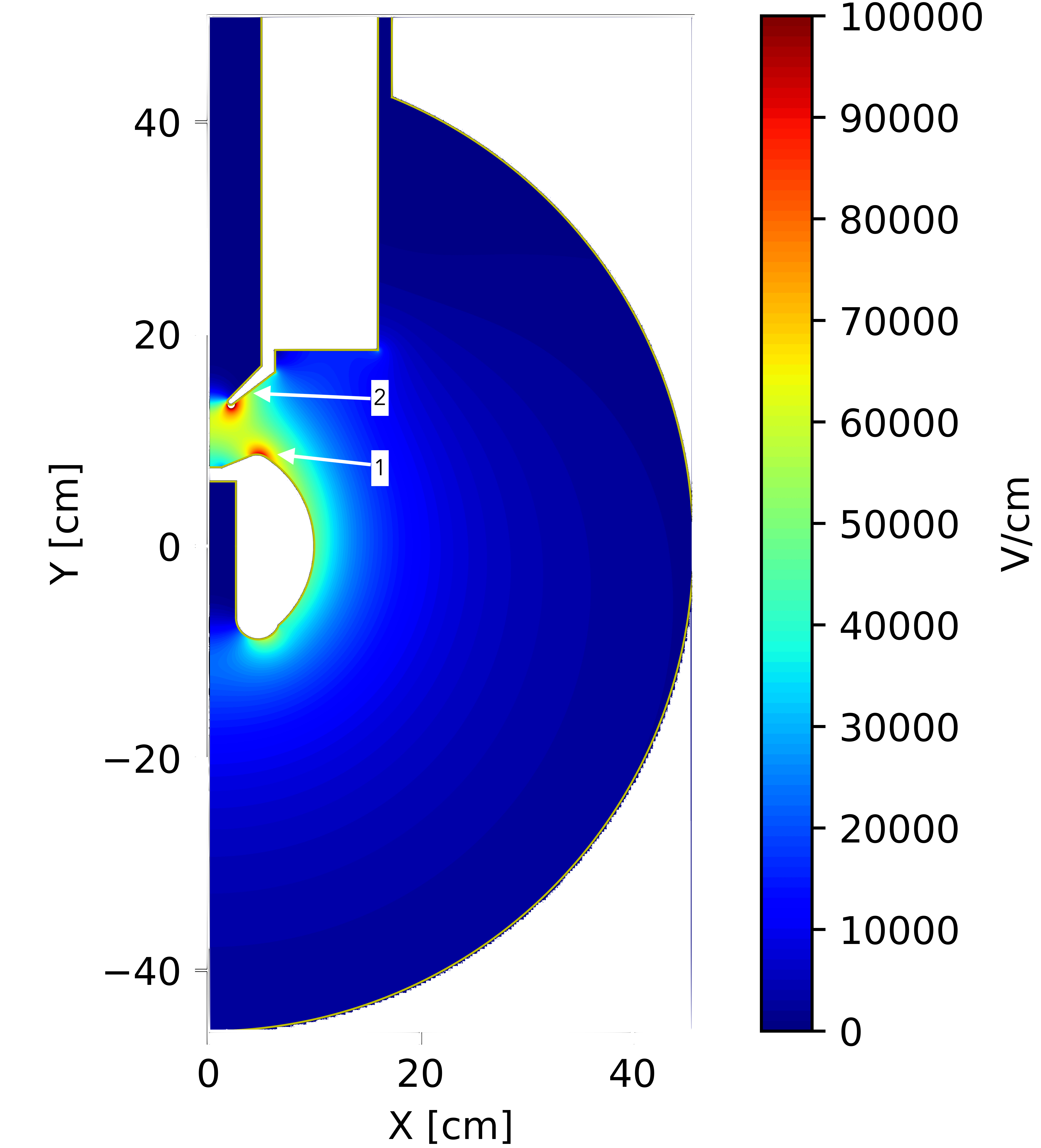}}
    \hfill
    \subfloat[\label{fig:feedthru}The field gradient on the HV ceramic feethrough.]
    {\includegraphics[width=1\columnwidth]{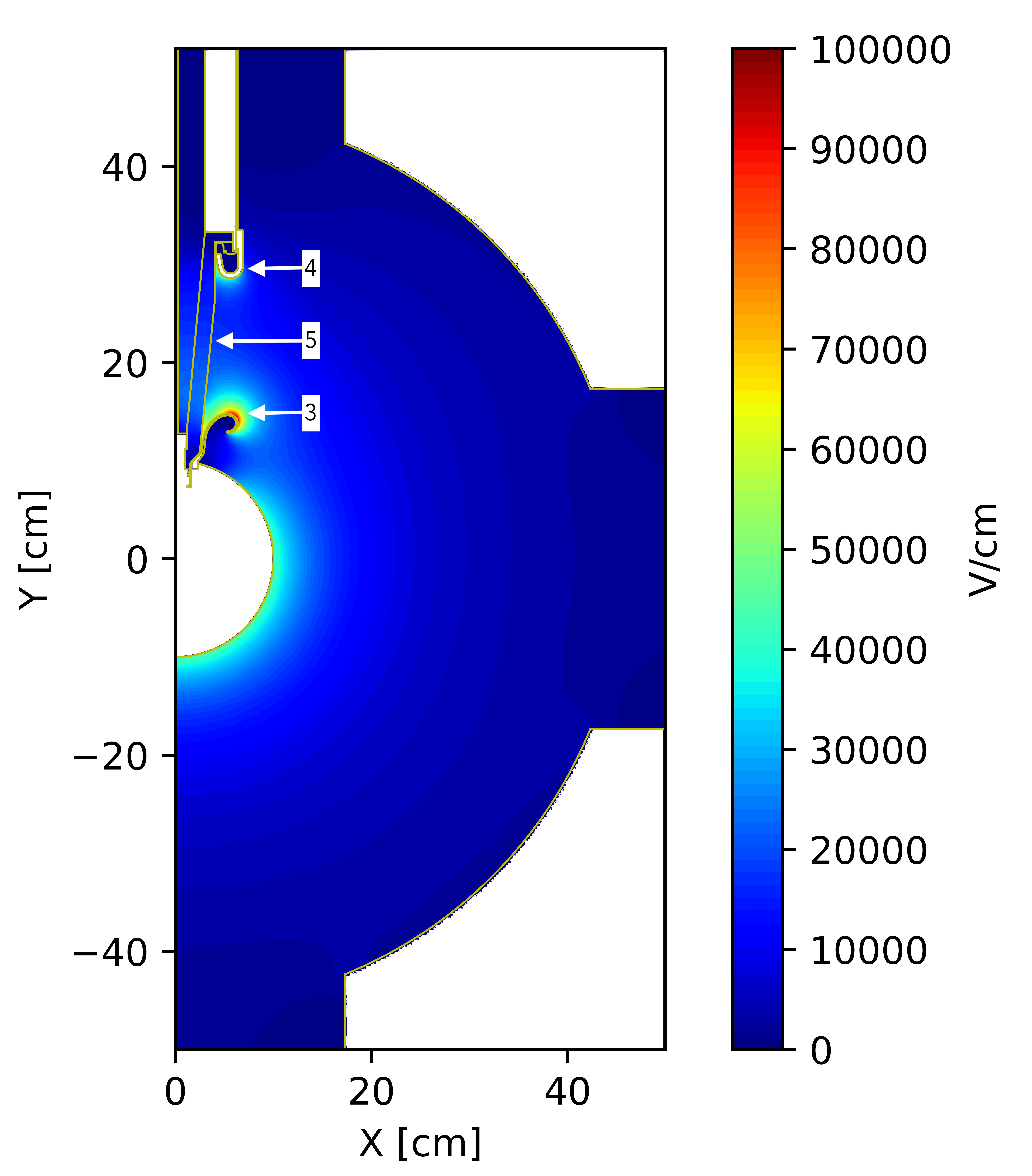}}
    \caption{The field gradient of the HVDC gun. The main high gradient locations are labeled: 1- Pierce nose; 2-anode; 3- high voltage; 4- grounded trip-point shed ; 5- ceramic}
\label{fig:2d_map}    
\end{figure}

The triple-point is the ceramic-vacuum-metal interaction point that usually has a local maximum field gradient. Two triple-point sheds were optimized and designed to minimize the gradient on the triple-point, and generate a uniform field gradient along the feedthrough to avoid electron trapping. The transverse size of the triple-point shed is determined by the transverse kick that it imparts on the electron beam. In our design, the deviation of the beam's trajectory from the cathode to the first corrector coil is less than 0.5 mm. The high voltage section's geometry was designed using Poisson and Opera3D and the field gradient is shown in figure \ref{fig:2d_map} \cite{Opera,poisson}.\par
At 350 kV gap voltage, the maximum gradient of 9.4\,MV/m occurs at the cathode Pierce shape nose(labeled 1) and the triple point shed(labeled 3) as shown in figure \ref{fig:2d_map}. The gradient at the center of the cathode surface is 3.8\,MV/m. The cathode linear field range is up to 9\,mm in radius. Therefore, We can offset the laser 9 mm or with a large laser size of 9\,mm in radius with increasing initial normalized emittance no more than 25\% \cite{omer19}.

We combined the mechanical polishing and SRF particulate cleaning process for our high voltage components. The stainless steel electrode and triple-point sheds were polished mechanically using a tumbler for an hour, following the method developed by JLab \cite{jlabbarrel}. The titanium anode was mechanically polished at a vendor using alumina powder. Then, the electrode, triple point sheds, anode, and the ceramic were ultra-sonic cleaned to remove residual surface contaminants and polishing media. Then High Pressure Rinsing (HPR) was performed on all metal parts with 1400 psi for 3 hours and 300 psi on ceramic for 3 hours, then dry them in a class 10 clean room overnight for pre-assembly. All the gun and beamline final installation was carried out in a class 10 clean room. \par

We use a Glassman high voltage power supply (HVPS) at a maximum of 400 kV.  It can provide a maximum current of 6 mA at this voltage.  To avoid the use of environmentally unfriendly SF6 or oil, the power supply is in the air within a grounded cage.  This places the HVPS about five meters from the gun. 
If there is an arc during the beam operation in the DC gap, the stored energy in the cable (which at 50 pF/ft is about 46 Joules) will damage the polished electrode. To avoid this from happening, we used a semiconducting layer as the shield of the cable with a resistivity of $10^{10}$ ohm-cm.  The resistivity is low enough to keep the cable from building charge but high enough to limit the delivery of stored energy from the cable in the DC gap. During conditioning, external 100 M$\Omega$ resistors were used in series, between the power supply and gun, to avoid damage to the polished electrode by a strong discharge.\par

\begin{figure*}[htp!]
\centering   
\includegraphics[width=0.9\textwidth]{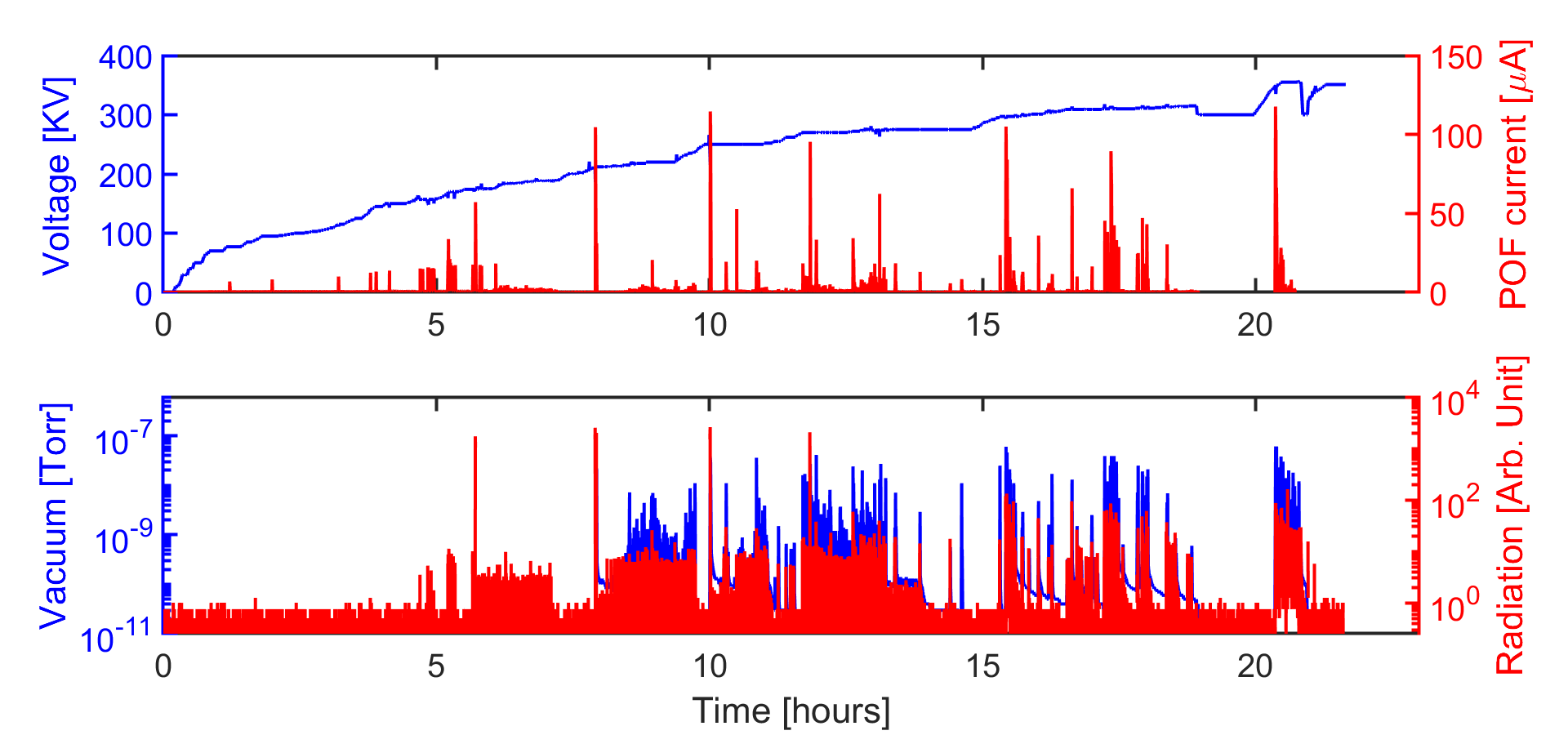}
\caption{\label{fig:hvpscondition}HVPS conditioning process to reach 350 kV.}
\end{figure*}

We installed a cathode puck with a GaAs wafer to condition the gun. The gun achieved 350 kV gap voltage after about 21 hours of conditioning. A few field emitters were found during the voltage ramp up, however, none of them were so called "hard barriers" that required more than 5-6 hours of vacuum conditioning. The trip limit for the power supply was set to 300 $\mu$A, which was never reached during the conditioning process. A Power over Fiber (POF) system was used to measure the current in the DC gap during gun conditioning. The maximum discharge we observed via POF measurement is about 120 $\mu$A, which occurred at 210 kV and 350 kV, and generated $10^{-8}$ Torr vacuum spikes. We did not have to use any inert gas, such as Helium or Krypton, for this conditioning process to achieve the design voltage. Vacuum excursions up to high $10^{-8}$ Torr were observed during the conditioning process, however, the vacuum in the gun could recover within the next few minutes after such excursion. We stopped to pursue even higher voltage because the electrode maximum gradient already reached 10 MV/m. Four Geiger counters were placed around the gun during the conditioning process. After the gun was conditioned up to 350 kV, the Geiger counters and field emission current showed background noise, and the vacuum got into $10^{-12}$ Torr scale. Figure \ref{fig:hvpscondition} shows the overall conditioning time along with the current, vacuum, and averaged radiation levels in the gun. Our total high voltage conditioning duration is much shorter than other comparable HVDC guns, indicating our new post treatment and clean processing might have played an important role \cite{Gu2020a,carlos19,Nishimori19}.

A (O,Cs) activated GaAs cathode was found to be field emission free up to 325 kV, compared to an un-activated cathode that can reach 350 kV. To avoid any chance of field emission during operation, we operated the gun at 300 kV with a sufficient safety margin. The field emission current at this voltage was lower than the noise, checked by the electrometer connected to the anode and current measurement from the POF system. We did not observe any QE drop overnight when the cathode was at 300 kV voltage without beam. 300 kV operational voltage for this polarized gun is already higher than any other, existing or retired, polarized electron sources \cite{grame2007,Nagoya2003}.


\section{Cathode heat transfer}
The gun was originally designed for mA scale average current operation \cite{erdong18}. Because the EIC changed collider scheme and electron source project changed the scope, our laser, machine protection system, and beam stop limited us to pursue average current beyond 70 $\mu$A. However, in the gun design and fabrication stage, we already implemented a cathode cooling system to avoid the cathode overheating by the incident laser power in high current operation \cite{erdong19}.\par

The gun side of the cable end is a modified, spring-loaded, R30 type plug. This modification was made by machining a pair of spiral grooves in the plug to accommodate the flow of Fluorinert (FC72) to the tip of the plug and back. The spring-loading provides good contact between the plug and ceramic feedthrough to maintain FC72 flow in the cooling channel.  A flow rate of greater than 0.5 gallons per minute is used, which does not result in cavitation. A chiller is used to maintain the FC72 fluid at room temperature. This specialized cable, inserted into the R30 ceramic feedthrough, was successfully tested at the Dielectric Sciences facility at up to 415 kV.\par
FC72, due to its a low viscosity, is known to be a fluid that is difficult to seal. MIT-Bates developed a cooling channel mechanism to cool the cathode, however, they faced a major problem of FC72 leaking into the vacuum system \cite{TSENTALOVICH2019,MITleak}. To assure there were no leaks into our vacuum system, we used continuous flow paths without decouplable mechanical joints in the FC72 flow path. Our RGA did not pick up any characterizing signal of FC72, showing that we had no leak into the system.\par
After assembling the cable plug into the ceramic and filling up the FC72 in the cooling loop, we occasionally observed air bubbles coming out from the plug due to gas trapped in the non-uniformity of the grease coating. This process will be mitigated by itself by redistributing the grease in about 12 hours of continuous circulation until there are no bubbles in the tube. Cameras are set up to monitor the flow paths for bubbles. We also found a continuous FC72 flow can dissolve the silicon grease and can leave grease residue in the return tube. The grease residue will change the cooling flow rate and results in degrading the cooling capability, therefore re-coating grease on the plug and cleaning the cooling loop have to be carried out every 3 months.
This setup has successfully operated the continuous FC72 flow more than a year without any leaks. We did not observe any failure in more than 500 hours of operation at high voltage.
A preliminary heat transfer test was carried out using this setup. The cathode temperature was seen to rise up to $60 \degree$C when 10 Watts of power was applied on the surface of the cathode without any cooling. Once the FC72 flow was implemented through the cooling channel, the cathode temperature dropped to $15 \degree$C. In principle, the $\rm K_2CsSb$ photocathode can survive at this temperature when 100 mA of average current is being generated, with laser power less than 10 Watts if the QE is above 2.3\% \cite{Gaowei2019}. Therefore, this cathode cooling setup can be applied in any high brightness high current electron source to be used for future scientific user facilities, such as electron cooling or energy recovery Linac based light sources \cite{coolingWang2021,eicbenson20,erleuv2021}.\par
\section{Gun test beamline}
\begin{figure}[h!]
    \centering
    \includegraphics[width=\linewidth]{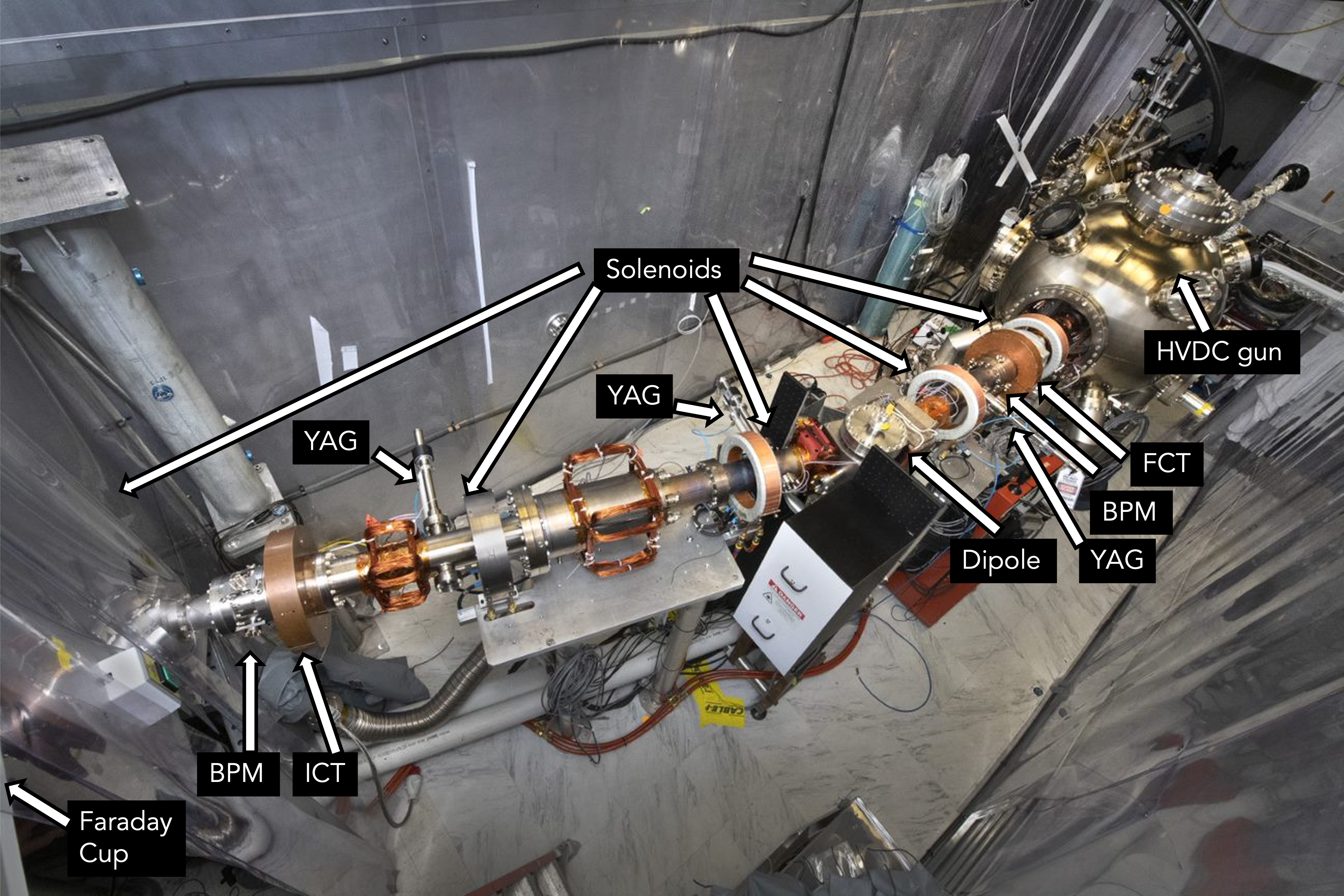}
    \caption{The HVDC gun test beamline. All the diagnostics are labeled.}
    \label{fig:layout}
\end{figure}

We developed a gun test beamline, for this gun, as shown in figure \ref{fig:layout}. Five solenoids maintain the beam envelop until the beam is delivered to the beam dump. A $16 \degree$ bending dipole is placed after the 2nd solenoid. The dipole magnet prevents the charged particles from downstream tracing back to the gun and also can be used for measuring the beam energy. The circularly polarized laser is delivered to the cathode, normal to the cathode surface, through the dipole chamber. The normal incidence of laser to the cathode surface is necessary to get high polarization. \par
One fast current transformer (FCT) and one integrated current transformer (ICT) are used to measure the bunch length and bunch charge. Four yttrium aluminum garnet(YAG) beam profile viewers are installed in the beamline to measure the beam transverse profile from the gun to the beam dump. The first YAG crystal has a 9 mm diameter aperture in the center, therefore the incident laser can pass through the viewer without scattering. It also can be used to measure beam halo. Two four-button beam position monitors (BPM) are used to monitor the beam position before and after the dipole during the operation. The aperture of the beam pipe is 10 cm in diameter which is 10 times the maximum RMS beam size for 5A peak current. The space charge beam tracking shows the beam halo can be well accommodated until it reaches the beam dump \cite{erdong19}. The planned differential pumping stages are not installed at this moment due to the change in the project scope.

\section{Gun and beam line Vacuum }
\begin{figure}[h]
\centering
\includegraphics[width=0.35\textwidth]{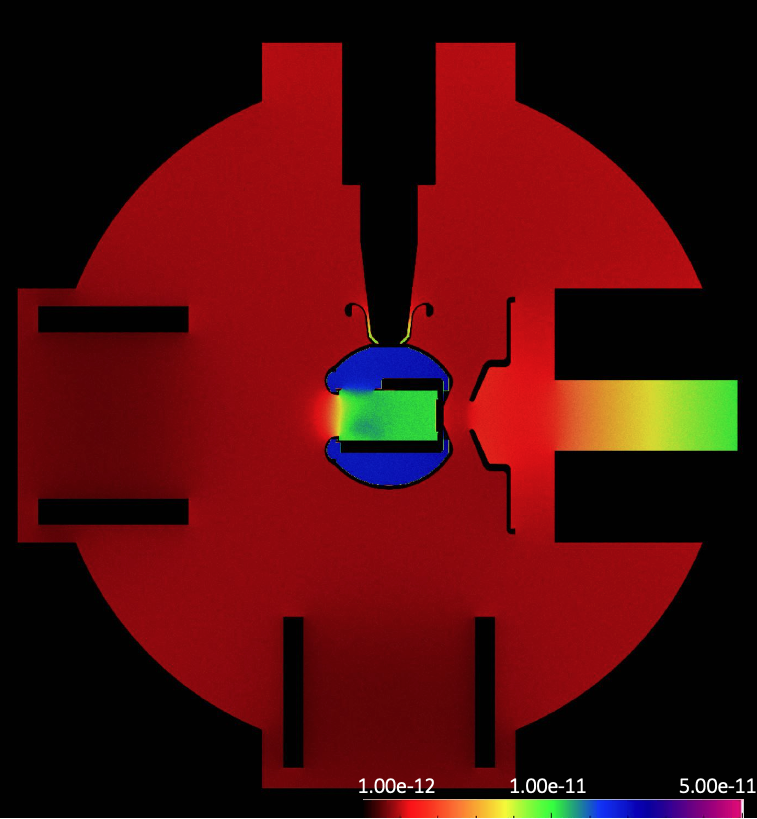}
\caption{The dynamic vacuum distribution in the gun chamber. In the simulation, we assume the Stainless Steel outgassing rate is $5\times 10^{-13} mbar\cdot L/s/cm^2$. The beam dump pressure was assumed to be $1 \times 10^{-8}$ $mbar$ for this simulation.}
\label{fig:vac_map}
\end{figure}

For the polarized electron source and beamline, a $10^{-12}$ Torr scale vacuum is required. In our system, the pumps in the gun area combine a sputtering ion pump(SIP) and 8000 \,L/s Non Evaporation Getters(NEGs) with ZAO material \cite{saes}. We performed a Monte-Carlo simulation using Molflow+ to estimate the pressure on the cathode surface and optimize the gun vacuum design \cite{molflow}. Figure \ref{fig:vac_map} shows the pressure distribution in the chamber, assuming beam induced outgassing at the beam dump. The beamline dynamic vacuum is typically worse than the pressure in the gun. To prevent the backstreaming molecules from reaching the cathode through the anode aperture, we designed a gap between the anode and vacuum chamber to provide extra vacuum conductivity. This will help in intercepting the gas coming from downstream before it reaches the cathode. \par
The choice of gun material and post-processing are crucial to achieving Ultra-high vacuum and high voltage. We chose the vacuum remelt cross forged 316 LN as our HV electrode material because it can achieve good smoothness due to its fine crystal. It was hydro-formed and fired up to $930^{\circ}$C for 4 hours to degas the trapped hydrogen in the material. 
The gun chamber and all the gun flanges were vacuum fired up to $930 \degree$C for 4 hours. Then the knife edges were machined. The thickness of the various stainless steel components, as applicable, was reduced to 5 mm to minimize the gas load after vacuum firing.
The anode is made out of Titanium due to its low atomic number. This will minimize the generation of X-rays when the field emission electron beam stops on the anode during gun conditioning. We also use a larger anode size, 32 cm in diameter, compared to other existing guns. The field emission from the highest gradient position, as shown in \ref{fig:2d_map}, will stop on the Titanium anode, other than the stainless steel chamber to further reduce X-rays. The anode is mounted on three electrically insulated actuators which provides flexibility to fine tune the anode position (in both longitudinal and transverse directions) and to be able to bias or use as a current pickup. The longitudinal movement can be used to optimize the voltage and optics in operation, while the transverse movements can be used for further extending cathode lifetime as proposed in the \cite{omer19}. In this paper, we describe the experiments with the anode positioned at the center of the cathode, with gap distance as shown in table \ref{tab:gun_parameter}.
The entire gun assembly was oven baked up to $350^{\circ}$C for 6 days for final outgassing.\par
The beam dump is 4.5 meters away from the HVDC gun. The first 0.5 m of beam pipe connected to the gun is coated with NEG material to further reduce the gas from the downstream beamline from reaching the gun. 
After activating all NEG pumps, the baseline vacuum was $7 \times 10^{-12}$ measured by the Bent Belt Beam gauge(3BG) gauge \cite{3bg2010}. 
During the beam test, we turned off the 3BG gauge because its hot filament will generate photons, which will in turn generate unwanted electron beams. The RGA measurements done at the end of the bake shows all gas species' partial pressure to be lower than $10^{-13}$(RGA noise level) Torr, except the hydrogen partial pressure at $4\times10^{-12}$ Torr. The dynamic vacuum of the gun, beam line, and beam dump will be discussed in detail in the next section.

\section{High intensity beam generation}
Due to the unavailability of superlattice GaAs in the market, we used bulk GaAs cathodes for our beam tests. The GaAs cathodes were activated using yo-yo process with Cs and O$_2$. The initial QE at 785 nm is about 5\%. It was transferred to the storage chamber first, then inserted into the HVDC gun. All the beam tests were carried out with gun voltage at 300 kV. Two pairs of Helmholtz coils were used to compensate for the earth's magnetic field. The laser repetition rate can be varied from 1 Hz up to 10kHz. The central wavelength for this laser is 785 nm with a frequency bandwidth of 1.3 nm.  In our experiments, we used a pulse length of 1.64 ns FWHM. The longitudinal pulse shape is flattop and the transverse profile is truncated Gaussian distribution. 
For 7.5 nC bunch charge, the beam size is 2 mm RMS in diameter right before it's entry into the beam dump. This measurement, along with radiation and vacuum levels on the beam line, indicates no beam loss in the beam transport.\par
\begin{figure}[h!]
    \centering
    \includegraphics[width=\linewidth]{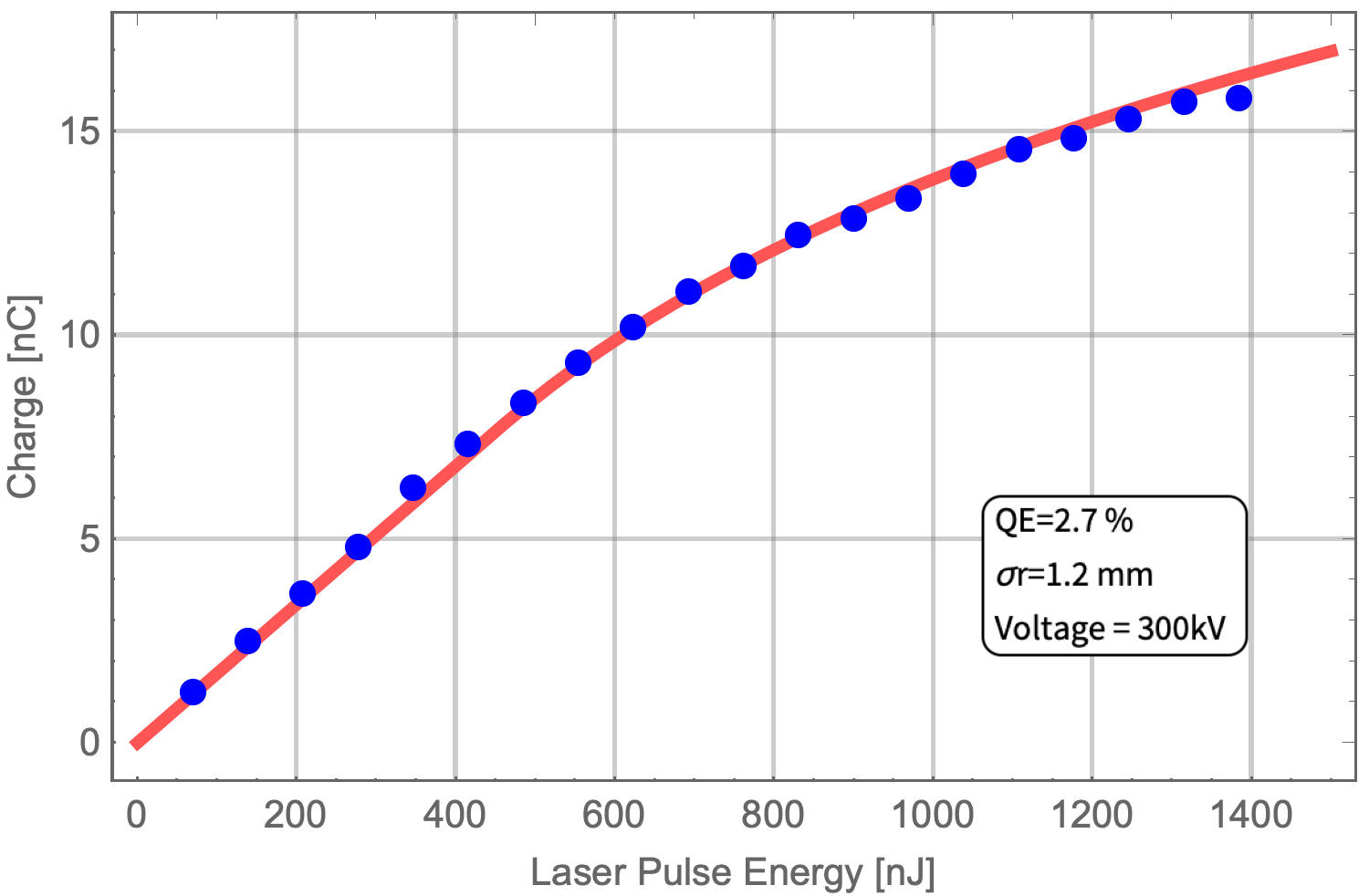}
    \caption{Space charge limit for the EIC polarized gun. Blue dots are measurement and the the red curve is the fitted curve.}
    \label{fig:SC}
\end{figure}

We measured the bunch charge as a function of laser pulse energy to identify the space charge limit for this gun. The cathode QE was 2.7\% during this measurement. The laser spot size on the cathode is 6 mm in diameter with a Gaussian sigma of 1.2 mm. The results are shown in figure \ref{fig:SC}. We fitted the data with 2D space charge limit model which shows the space charge limit occurs at 12 nC at this beam size \cite{HARTMAN1994219}. Note that we defined the space charge limit to be the point at which the peak of the current density Gaussian distribution starts to saturate, not as the maximum extractable bunch charge. For example: even though our gun could deliver beyond 16 nC with a 6 mm laser spot size diameter, the space charge limit is at 12 nC according to the above mentioned definition that we adopted. Our cathode size is 26 mm in diameter, therefore, a much higher space charge limit could be achieved if we use a larger laser spot size. The EIC nominal charge 7 nC is well below the space charge limit for a 6 mm spot size.  \par 

For cathode lifetime tests, we generated 7.5 nC bunches from an area 6 mm in diameter. The beam was transported about 4.5 meters downstream to the beam dump without any measurable loss and the bunch charge was measured using the ICT at the end of the beamline. We performed gun beam tests for more than 100 hours at various operational modes. we did not observe any HVPS trips during operation, showing the gun operation is very stable.  

\begin{figure}[t!]
    \centering
    \includegraphics[width=\linewidth]{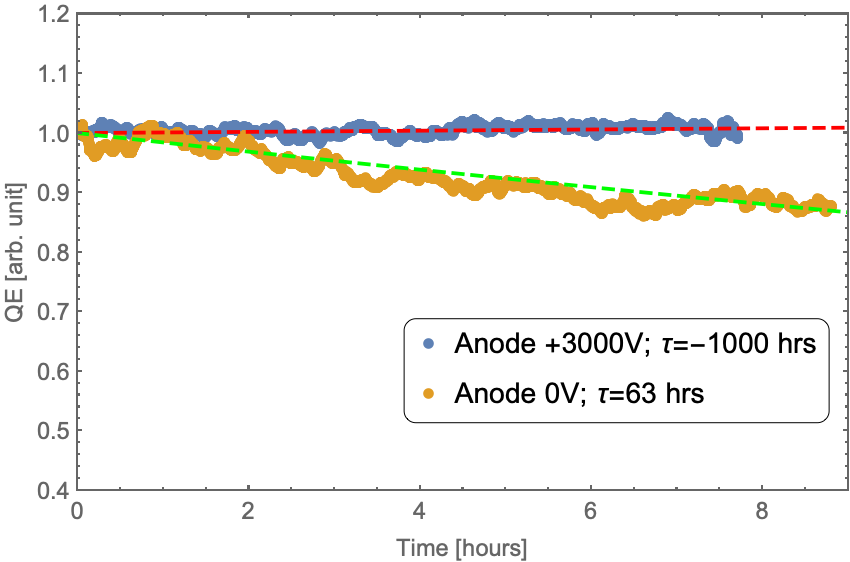}
    \caption{Comparison of QE decay between unbiased anode and biased anode operation. The peak and average current in both cases were the same - 3.5 A and 37.5 uA respectively. The $\tau$ is the decay time using exponential fitting. The negative $\tau$ means the QE had slightly increased in 8 hours.}
    \label{fig:QE_tets}
\end{figure}

\begin{figure*}
\centering
\subfloat[\label{fig:5a}]
{\includegraphics[width=0.3\linewidth]{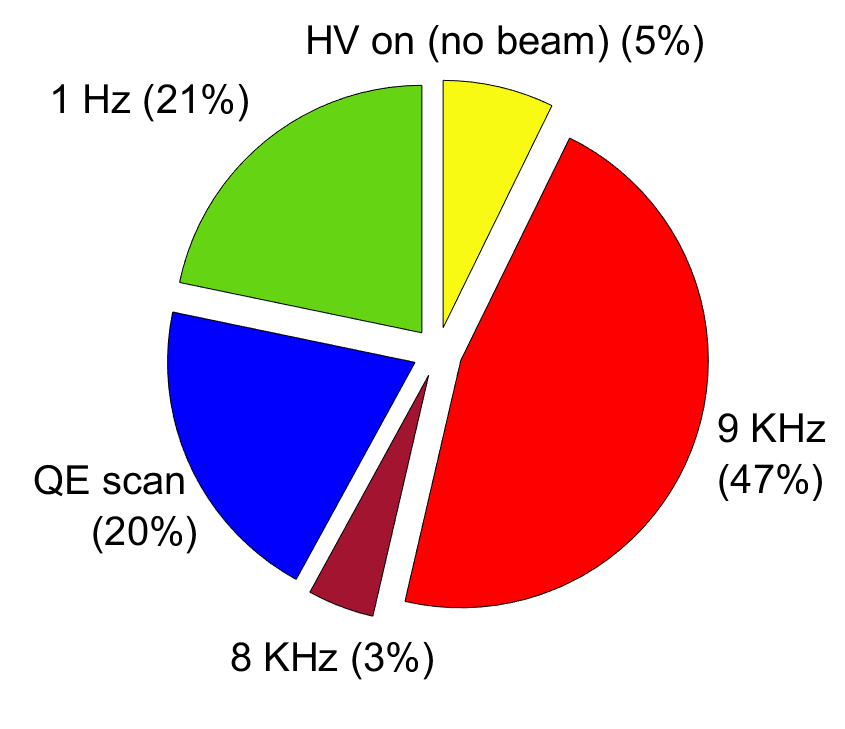}}
\hfill
\subfloat[\label{fig:5d}Before beam]
{\includegraphics[width =0.3\textwidth]{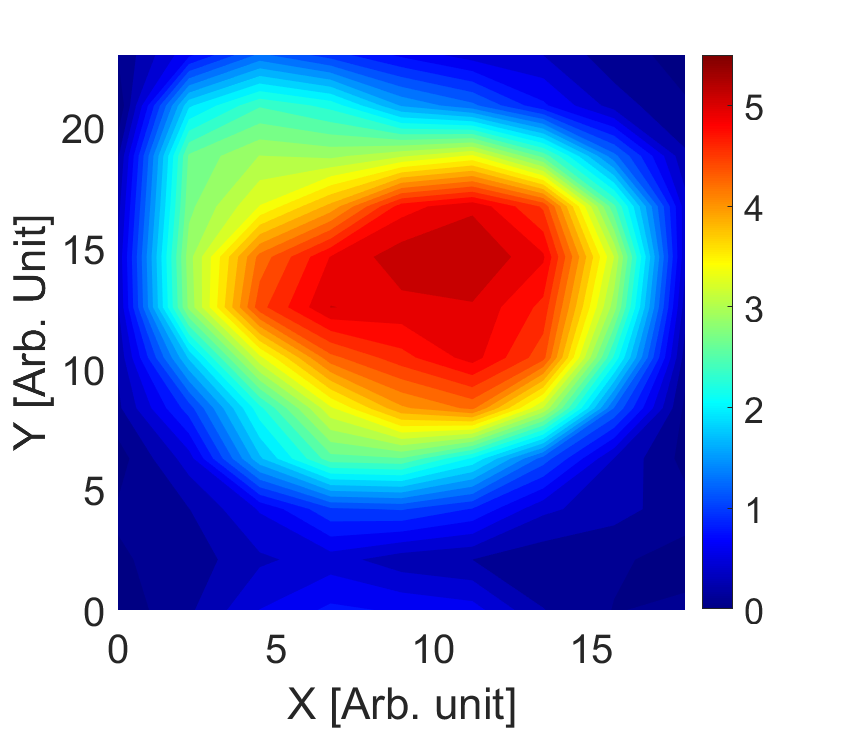}}
\hfill
\subfloat[\label{fig:5f}After beam]
{\includegraphics[width =0.3\textwidth]{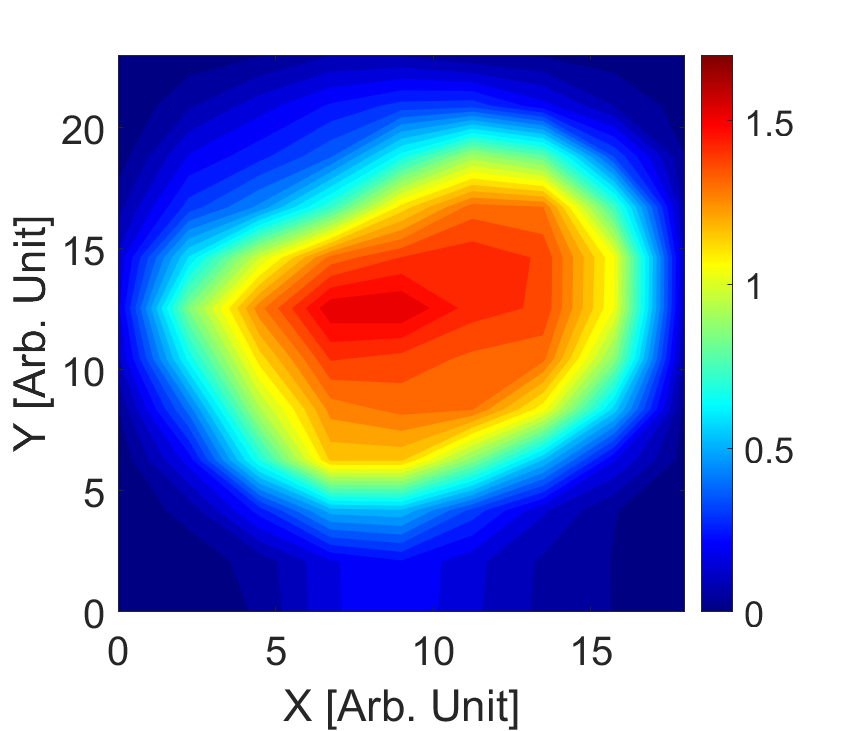}}
\caption{Left: Fractional amount of time spent on a sample cathode for different operation modes. The percents in the bracket indicate the percentage of overall operation time from this cathode. Middle: QE scan of the cathode before any beam was extracted. Right: QE scan after the cathode was used as per the left pie chart. The Colorbar represents QE in percent (\%).}
\label{fig:qescan}
\end{figure*}

Lifetime tests were performed for various average currents - 1.5 $\mu$A, 15 $\mu$A, 37.5 $\mu$A, and 67.5 $\mu$A - all orders of magnitude higher than the EIC requirement. The vacuum levels on the beamline and the beam dump were measured using hot filament ULVAC AxTRAN gauges. A slow feedback system for the laser pulse energy was employed, so that the bunch charge could be kept constant while adjusting laser pulse energy. The adjustment of the laser pulse energy occurred every 15 seconds if the bunch charge changed by 5\%.  For the 1.5 $\mu$A and 15 $\mu$A cases, no observable QE decay was seen after 15 hours of continuous operation. For the 37.5 $\mu$A operation, 10\% QE decay in 9 hours was observed when the anode was not biased.  For this operation mode, the beam stop vacuum rose up to $7 \times 10^{-10}$ Torr, whereas the beamline vacuum rose to $1.5\times 10^{-11}$ Torr. During all operations as stated so far, the cold cathode gauge in the gun vessel was below its measurement threshold. However, the cold cathode gauge can only be trustworthy down to $5\times 10^{-11}$ Torr. The gun vacuum pressure may increase to high $10^{-12}$ Torr to low $10^{-11}$ Torr scale, without the cold cathode gauge registering any reading.\par 

Ion back bombardment was observed and determined as the main QE decay when operating at or above 10s $\mu$A level of average currents in various facilities \cite{Grames2008,Wei2016}. In our particular case, we have to identify whether the QE drop was due to dynamic residual gas or ion back bombardment. Therefore, We repeated the exact same experiment with the same average current, this time biasing the anode to +3\,kV. Any ions generated from the beamline should be blocked by the biased anode. For the biased anode case, no observable QE decay was found over 8 hours of operations. Therefore, the ions generated in the beamline dominated the QE decay when the anode was not biased. Figure \ref{fig:QE_tets} shows the cathode QE during operation at 37.5 $\mu$A average current, comparing the biased and unbiased anode operation modes. \par

The beam line is not suitable for operating high average current polarized beam without differential pumping stations, with conduction limitations, between the beam dump and the gun to intercept the back streamed gas. 67.5 $\mu$A average current operation showed clear QE decay during operation even with the anode biased. During this test, the gun vacuum would rise to $1-2 \times 10^{-11}$ Torr as measured from the cold cathode gauge. The beam dump and beam line vacuum read $2\times 10^{-9}$ Torr and $4 \times 10^{-11}$ Torr respectively. This rise in vacuum is a direct result of the rise in beam stop and beam line vacuum. An increase in the gun vacuum will increase chemical poisoning and ion back bombardment in the DC gap. This phenomenon was clearly seen during the first beam test at 67.5 $\mu$A as 10 hours of operation showed about 20 $\%$ QE decay. At this point, the maximum average current and charge lifetime in this particular gun is limited by the beam dump out gassing. A short test beamline is usually not suitable to test a high current beam. In the collider, the beam dump will be kilometers away and the effects on the cathode will be negligible.  \par

We compared the QE map before and after 67.5 $\mu$A average current operation for 39 hours, as shown in figure \ref{fig:qescan} pie chart as 9 kHz operation. Figure \ref{fig:qescan} (left) shows different operation modes for a sample cathode in 785 nm laser. For this cathode, the majority of the operation was at 67.5 $\mu$A average current. The bunch charge was 7.5 nC for each mode, except for the QE scan. A 0.5 mm spot size was used during QE scan and extracted bunch charge was about 80 pC. The QE scans before and after the tests, as depicted by the pie chart, were shown in figure \ref{fig:qescan}. Comparing the before and after scans, it seems that for this particular run, we didn't observe the typical ion back bombardment causing a sharp QE drop at the center of the cathode. 
From previous experiments at various facilities and simulations, a signature of ion back bombardment dominated QE decay is generally seen in major QE loss at the electrostatic center of the cathode. In our case, the decay is rather uniform over the entire activation area rather than small deep trenches in certain spots. This result indicates that back chemical poisoning induced by back streamed gas could play a key role in degrading the cathode. \par
\begin{figure}[h!]
    \centering
    \includegraphics[width=\linewidth]{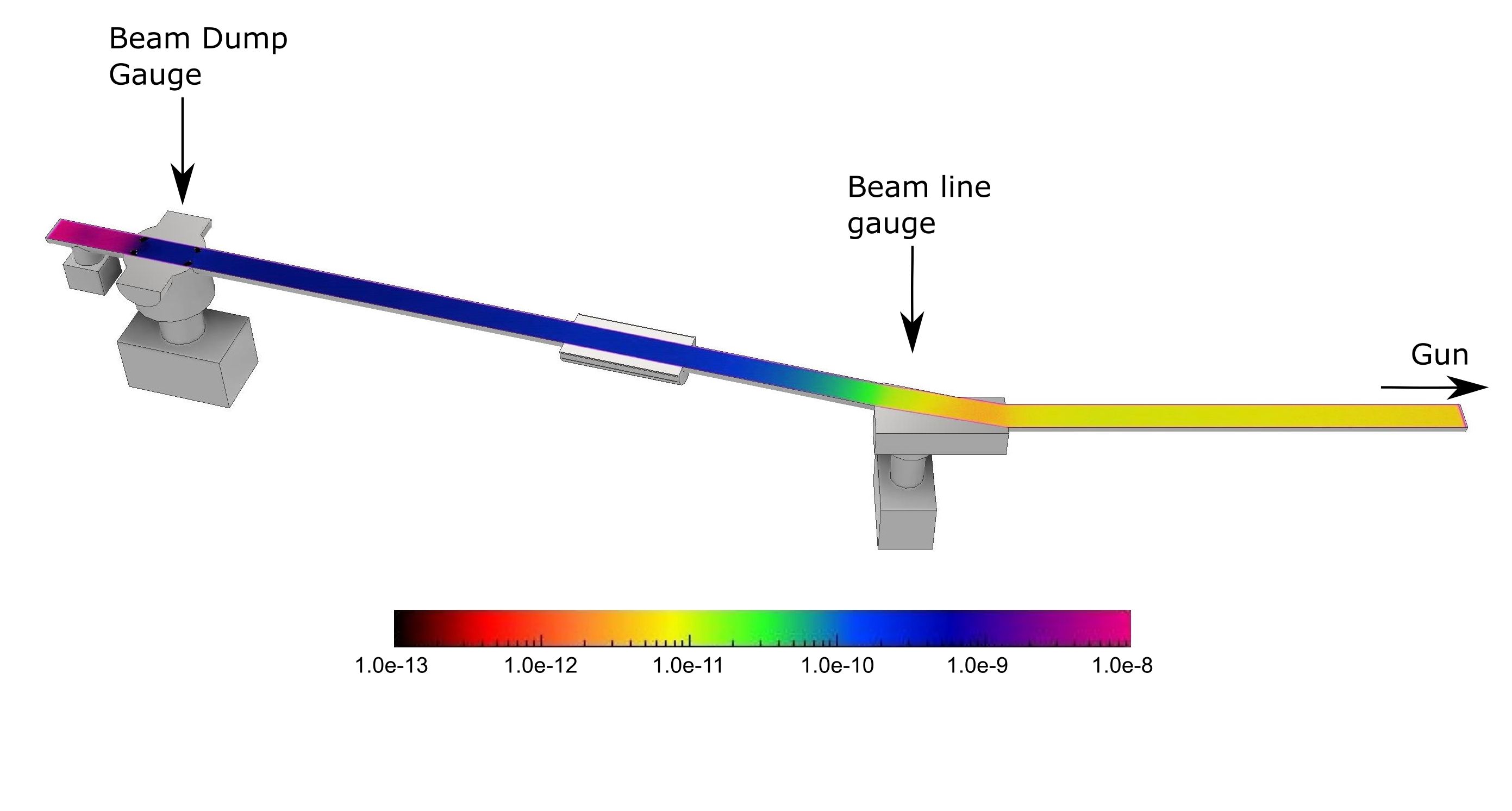}
   \caption{Molflow+ simulation of the beam line, assuming a $1\times 10^{-9}$ Torr vacuum in the beam stop.}
    \label{fig:molflowdynamic}
\end{figure}

We simulated the dynamic vacuum in the beam line using Molflow+. Assuming $1 \times 10^{-9}$ Torr vacuum in the beam stop, and with the existing pumping on the beam line as outlined in the previous sections, we found that the rise in the gun vacuum to $10^{-11}$ Torr scale is reasonable. The results are shown in figure \ref{fig:molflowdynamic}.\par

The EIC gun parameters require 0.0338 C charge to be delivered in a week. During the 37.5 $\mu$A continuous operation of 7.5 hours, with no observable QE decay, we delivered approximately 30 times of EIC's weekly charge requirement.
Therefore it is accurate to say that this gun has exceeded the EIC requirements by a substantial margin.
\begin{figure}[h!]
\centering
\includegraphics[width=0.92\linewidth]{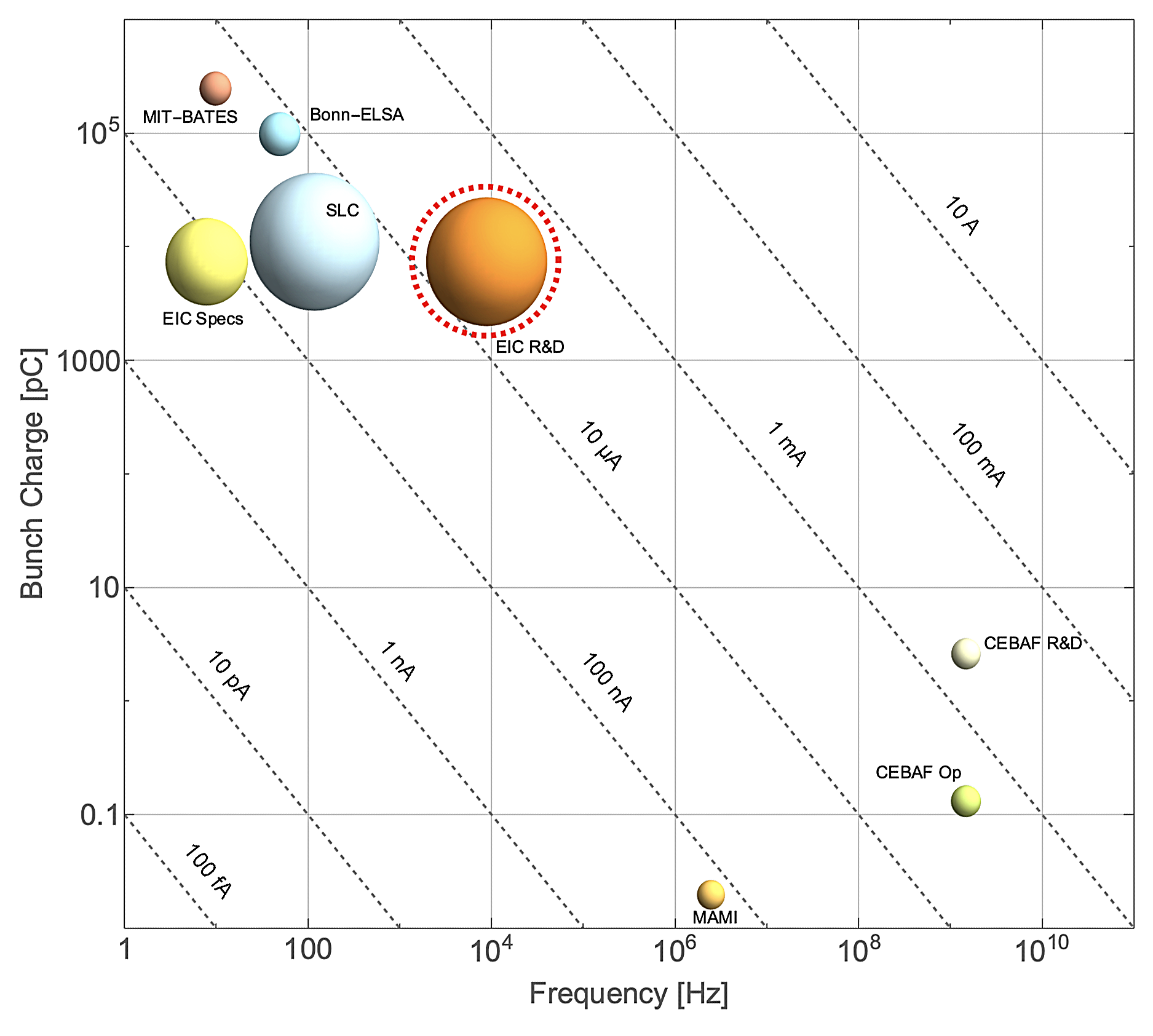}
\caption{Comparison of various gun from different facilities around the world. The slope line shows the average current contour level as labeled. The ball diameter is representative of the peak current of the gun. The red dash line at EIC $\rm R\&D$ shows the maximum achieved peak current of 8\,A. }
\label{fig:gun_review}
\end{figure}
Figure \ref{fig:gun_review} shows a comparison between various polarized guns around the world. Other operational guns, past and present, would either have a high bunch charge ($\>$ 1 nC) or high average current ($\>$10s $\mu$A) but not both. For example, CEBAF operation and R\&D gun could operate in a range between 100 $\mu$A to several mA level average current but with pC level bunch charge \cite{adderley2011cebaf}. On the other hand SLAC and MIT guns had higher bunch charge but with relatively low average current \cite{clendenin1995polarized}. This EIC gun is unique because it can operate in a regime where higher average currents (100's $\mu$A) can be delivered with bunch charges in 10 nC level.

\section{Conclusion}
We have designed and commissioned a HVDC polarized electron gun to meet the EIC polarized electron beam requirement. This gun employs various novel concepts, including a cathode cooling system which could be implemented in future high current electron sources. The gun was conditioned up to 350 KV without any field emission and was consistently operated at 300 KV. High bunch charge, up to 16 nC, the beam was generated from bulk GaAs photocathodes using a 785 nm laser. Gun performance, including operational lifetime, exceeds all EIC requirements. Lifetime experiments with up to 37.5 $\mu$A level average currents, with a biased anode, show no observable QE decay over 15 hours. At this point, the lifetime for 67.5 $\mu$A average current from this gun is limited by beam dump outgassing.

\section{Acknowledgement}
The authors would like to acknowledge Ronald Napoli for assembling the gun and beamline.
The authors would like to thank Matthew Poelker, Carlos Hernandez-Garcia and Joe Grames from Jefferson lab for useful discussions, and Don Bullard from JLab for polishing the high voltage electrode. The authors are thankful to Zachary Conway and the SRF group of Collider Accelerator Department for electrode cleaning process. The authors would like to acknowledge Charles Hetzel and NSLS II vacuum group at BNL for vacuum support. Finally, the authors would like acknowledge the following vendors with the design effort: Jerry A. Goldlust,Peter Dandridge - Dielectric Sciences;
Mike Ackeret and Jason Pietz - UHV Transfer Systems;
Tom Casey, Tom Beard,  Richard and Jed Bothell - Atlas Technologies;
Tom Bogdan - MDC.
Enrico Maccallini, and Paolo Manini - SAES Group.

The work is supported by Brookhaven Science Associates, LLC under Contract No. DE-SC0012704 with the U.S. Department of Energy.

\bibliography{HVDC}

\end{document}